\documentclass[10pt]{article}

\usepackage{amsmath}
\usepackage{amssymb}
\usepackage{graphicx}
\usepackage{setspace}
\usepackage{cite}
\usepackage{hyperref}
\usepackage{url}
\usepackage{color}

\usepackage[labelfont=bf,labelsep=period,justification=raggedright]{caption}

\makeatletter
\renewcommand{\@biblabel}[1]{\quad#1.}
\makeatother

\usepackage{xcolor}

\date{}

\begin{document}


\begin{flushleft} {\Large \textbf{Geography and similarity of regional cuisines in China} }
\\
Yu-Xiao Zhu$^{1,2,3}$,
Junming Huang$^{4}$,
Zi-Ke Zhang$^{5}$,
Qian-Ming Zhang$^{1,3}$,
Tao Zhou$^{1,3}$,
Yong-Yeol Ahn$^{2,\ast}$
\\
\bf{1} Web Sciences Center, School of Computer Science and Engineering,
University of Electronic Science and Technology of China, Chengdu, Sichuan, People's Republic of China
\\
\bf{2} School of Informatics and Computing, Indiana University, Bloomington 47408, United States of America
\\
\bf{3} Beijing Computational Science Research Center, Beijing 100084, People's Republic of China
\\
\bf{4} Institute of Computing Technology, Chinese Academy of Sciences, Beijing 100190, People's Republic of China
\\
\bf{5} Institute of Information Economy, Alibaba Business College, Hangzhou Normal University, Hangzhou, Zhejiang,  People's Republic of China
\\
$\ast$ Corresponding Author: yyahn@indiana.edu
\end{flushleft}

\section*{Abstract}

Food occupies a central position in every culture and it is therefore of great interest
to understand the evolution of food culture. The advent of the World Wide Web
and online recipe repositories has begun to provide unprecedented opportunities
for data-driven, quantitative study of food culture. Here we harness an online
database documenting recipes from various Chinese regional cuisines and
investigate the similarity of regional cuisines in terms of geography and
climate. We found that the geographical proximity, rather than
climate proximity is a crucial factor that determines the similarity of regional cuisines.
We develop a model of regional cuisine evolution
that provides helpful clues to understand the evolution of
cuisines and cultures.


\section{Introduction}

The most essential need for all living organisms is energy, which is usually obtained by
consuming food. So it is not hard to imagine why food affects all aspects of human
life and culture~\cite{foodandcuture,audrey1932,raymond1934}. Food has been
studied in depth by  many disciplines including
history~\cite{history1,history2,history3,history4},
sociology~\cite{sociology1,sociology2,sociology3},
philosophy~\cite{philosophy1,philosophy2}, and literary
criticism~\cite{literary1}, etc. Understanding how food culture
evolves will have profound impact on numerous domains.
Despite its manifest
importance, few studies have taken quantitative, systematic approaches towards
food culture, mainly due to the scarcity of systematically collected databases.

Nevertheless, such an approach is promising. For example, one
pioneering study  revealed the connections between climate and the use of
spices through the manual digitization of a large number
of traditional recipes~\cite{spices}. Recent increases in online recipe
repositories have begun to allow for easier access to structured recipe
data~\cite{YY1,osameNJP,YY2}. Harnessing this opportunity,
we address the following questions about the evolution of regional
cuisines: (1) How does the similarity between regional cuisines scale with
geographical distance? (2) Is climate similarity the main factor determining
similarity? Climate obviously plays an important role
in shaping food culture, because it both limits the availability of
ingredients and affects the usage of spices~\cite{spices}. However, geographical proximity alone might drive
nearby cuisines close because of frequent communication and migration. To address these questions, we examine regional recipes in China --- the second
largest and the most populous country in the world, which is
home to billions of people with diverse cultural heritages. The landmass of
China spans North-South and East-West rather evenly, providing a good test bed
to study the effect of geography and climate.

Our investigation suggests that geographical distance alone plays a more important
role than climate. Based on our
results, we propose a model of cuisine evolution based on the copy-mutate
mechanism. The key idea is that nearby
regional cuisines tend to learn from each other either due to communication or
migration. We demonstrate that our simple model reproduces important
characteristics of the real data.

\section{Data and Methods}
\subsection{Data collection}
\begin{figure*}[htpb]
  \centering
  \includegraphics[width = 15.1cm,height=15.1cm]{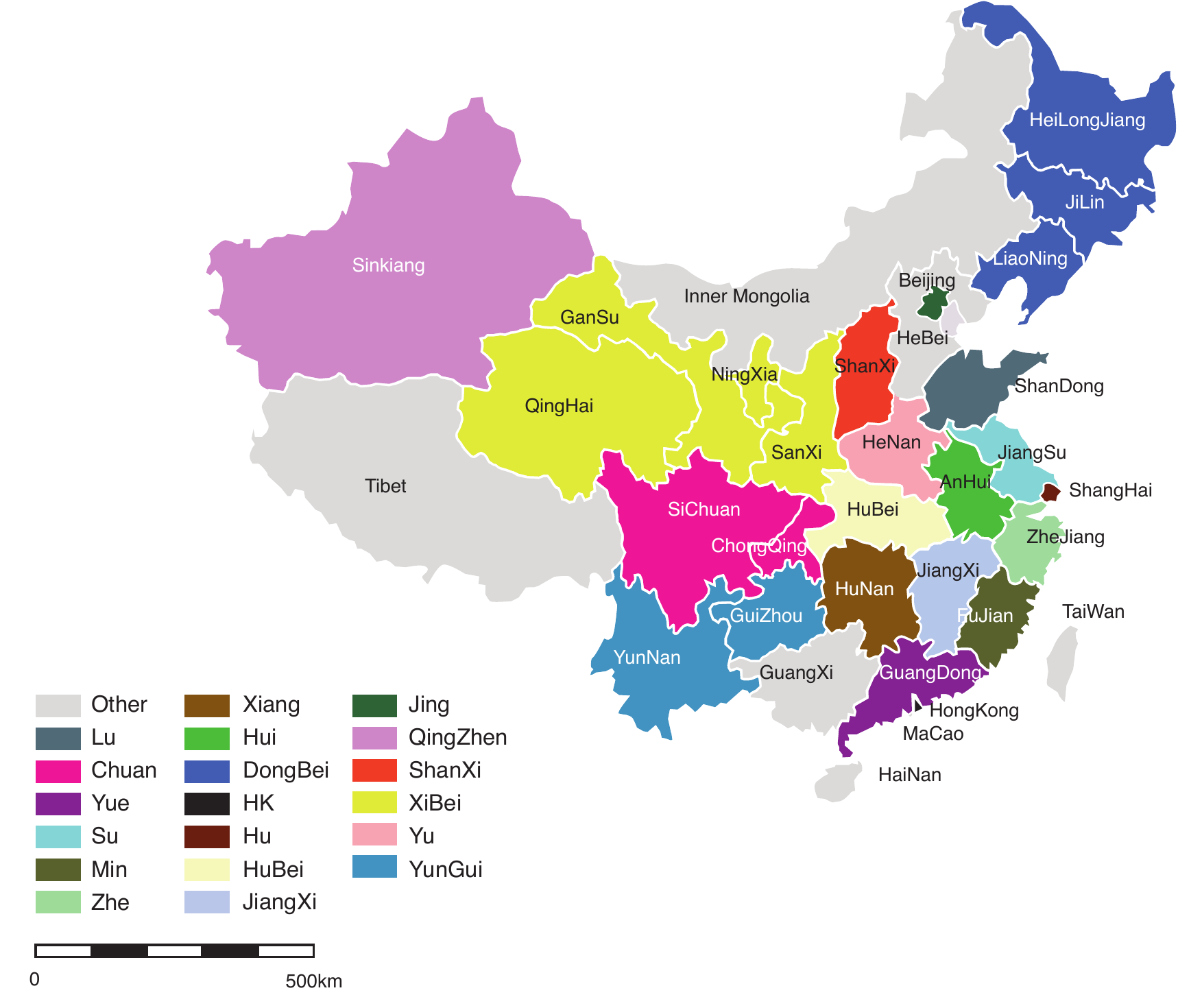}\\
  \caption{ {\bf Map of regional cuisines in China.}}\label{map}
\end{figure*}

\begin{figure}[htpb]
  \centering
  \includegraphics[width = 6.2cm,height=4.7cm]{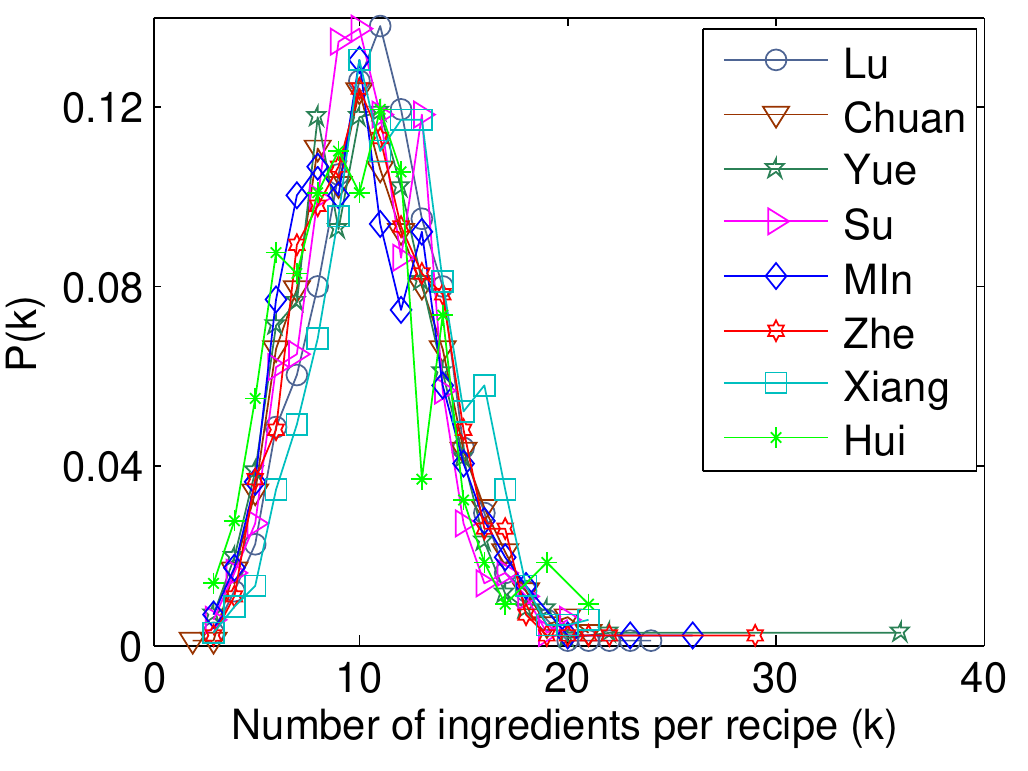}\\
  \caption{ {\bf Probability distribution of the number of ingredients per recipe.}
  All cuisines sow similar distributions, which have a peak around $10$. }
  \label{number_ingredients_per_recipe}
\end{figure}

In April 2012 we downloaded all the recipes from the Chinese recipe website \emph{Meishijie}\footnote{http://www.meishij.net/},
which categorizes all recipes into $20$ regional cuisines. Figure~\ref{map} shows a map of China, annotated with the names of
provinces and cuisines, where the colors represent cuisines. Multiple
provinces may belong to the same cuisine. Each recipe has the following
properties: (i) a cuisine (each recipe belongs to only one cuisine); (ii) a
list of ingredients; and (iii) a cooking method. We manually consolidated
synonymous ingredient names and removed the
recipes that do not have cooking method or have too few (less than three)
ingredients. There are 8,498 recipes and 2,911 ingredients in the cleaned
dataset. Basic statistics of the dataset are reported in
Table~\ref{cuisine}. Figure~\ref{number_ingredients_per_recipe} shows
that the number of ingredients per recipe contains is similarly
distributed across all regions (the mode is around 10).

\begin{table*}[htpb]
  \centering
  \caption{The basic statistics of regional cuisines in China. }
  {\begin{tabular}{cccccccccccccccccccccc}
  \hline\hline Regional Cuisine& $N_{r}$ & $N_{i}$ & $\hat{N_{i}}$ & $\langle K \rangle$ &Location\\
  \hline
  \textbf{Lu}&      1,066&  788&  64&  10.8452& ShanDong \\
  \textbf{Chuan}&   1,148&  877& 195&   10.4608& SiChuan, ChongQing \\
  \textbf{Yue}&      775&   900&  23&  10.2839& GuangDong  \\
  \textbf{Su}&      372&   573&  88& 10.3199& JiangSu \\
  \textbf{Min}&     468&   648&  69&  10.1880& FuJian \\
  \textbf{Zhe}&     460&   594&  14&  10.5978& ZheJiang \\
  \textbf{Xiang}&   691&   592&   7&  11.4660& HuNan \\
  \textbf{Hui}&     218&   442& 111&    9.8899& AnHui \\
  DongBei& 358&   458& 237&   10.0363& JiLin, HeiLongJiang, LiaoNing \\
  HK& 151&   367& 210&    9.4040& HongKong\\
  HuBei&    160&   266& 115&   10.7375& HuBei\\
  Hu&       744&   692& 117&    9.4274& ShangHai \\
  JiangXi& 143&   210&  87&   10.3147& JiangXi \\
  Jing&    606&   565&  74&  10.3614& BeiJing \\
  Other& 52&    171&  16&   9.5962& - \\
  QingZhen&   521&   426& 131&  10.7524& Sinkiang \\
  ShanXi&   125&   191&   5&  11.4720& ShanXi \\
  XiBei&   188&   338&  52&  10.8351& SanXi,GanSu, QingHai, NingXia \\
  Yu&  173&   291&  10&  10.6936& HeNan \\
  YunGui&   79&    184&  23&   8.8101& GuiZhou, YunNan\\
  \hline
  \textbf{All}&    8,498& 2911 &  - &  10.4399 &  - \\
  \hline\hline
  \end{tabular}
  \begin{flushleft} There are  $M=20$ different regional cuisines in total. The eight major regional cuisines, labeled in bold, are the most representative and typical cuisines in China. $N_{r}$: number of recipes. $N_{i}$: number of ingredients. $\hat{N_{i}}$: number of ingredients used only in the  cuisine. $\langle K \rangle$: average number of ingredients in a recipe. The last column reports the provinces where a regional cuisine originates in.
  \end{flushleft}
  \label{cuisine}}
\end{table*}

\begin{figure*}[htpb]
  \centering
  \includegraphics[width =7cm,height=5cm]{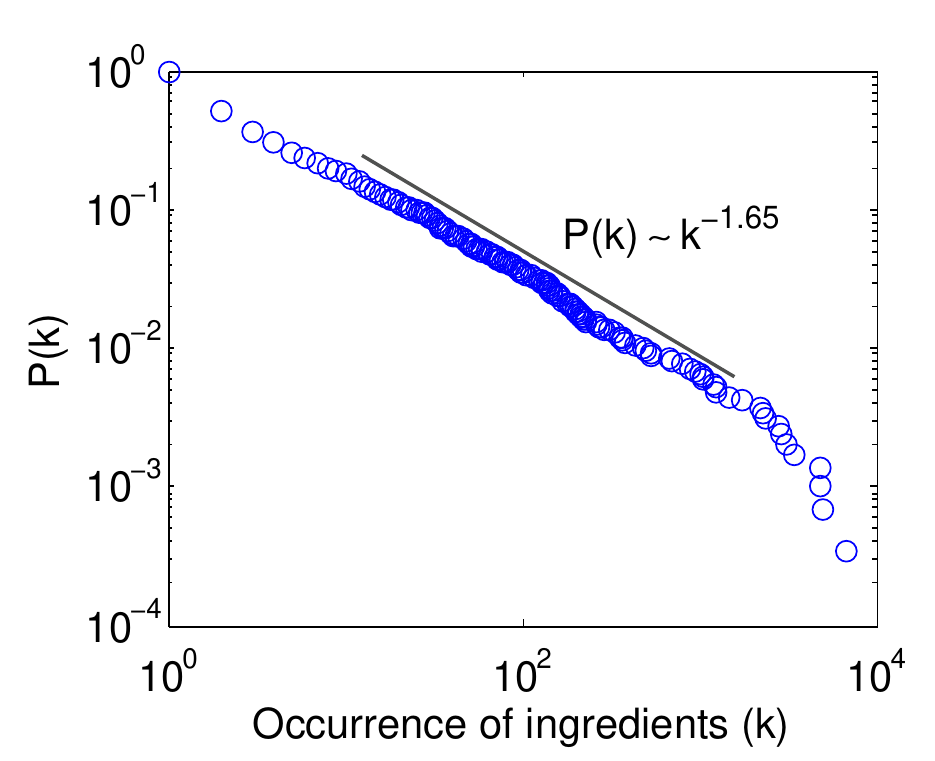}\\
  \caption{ {\bf Cumulative frequency distribution of ingredient usage.} The usage frequency is calculated using all recipes in our dataset. The exponent is obtained by the method in paper~\cite{powerlaw}. } \label{degree_ingredient}
\end{figure*}

\begin{figure*}[htpb]
  \centering
  \includegraphics[width =7cm,height=5cm]{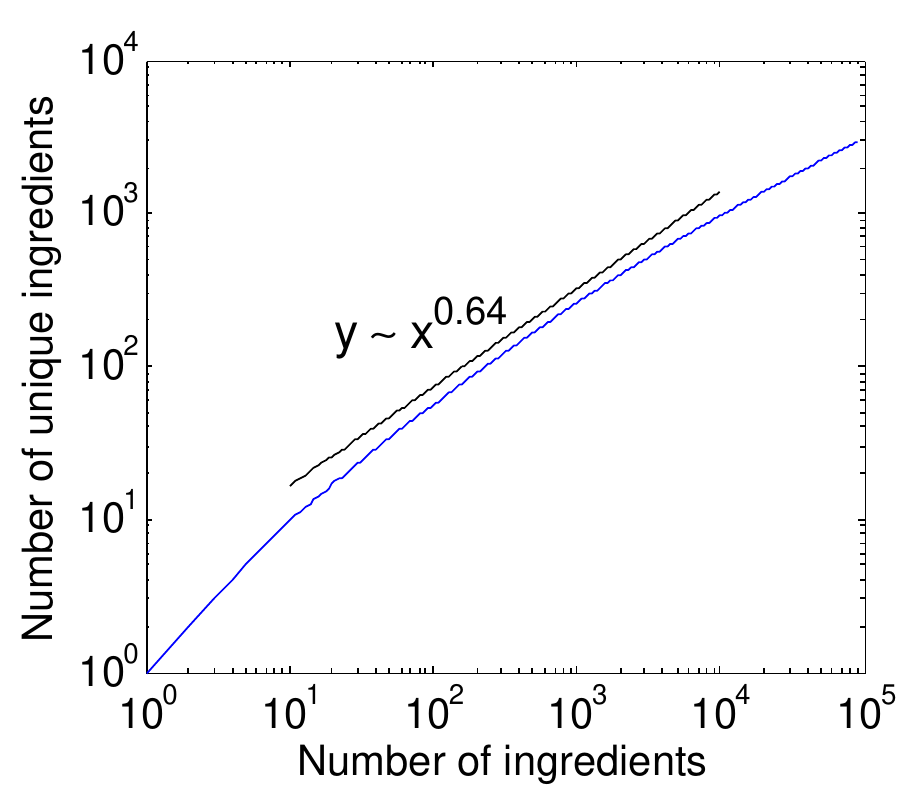} \\
  \caption{ {\bf The number of distinct ingredients discovered VS. the number of recipes scanned.} The plot (blue curve) approximately follows  Heap's law (black guideline). The graph is obtained by averaging $100$ implementations with independently random sequences of recipes. } \label{heap_law}
\end{figure*}

As previously observed, the ingredient usage frequency follows a skewed
distribution; only a small fraction of popular ingredients are widely used,
while many ingredients are observed in only a small number of recipes. As
shown in Fig.~\ref{degree_ingredient}, the frequency distribution of
ingredients follows a power-law~\cite{powerlaw-barabasi}
($\lambda \sim 1.65$ using method in paper~\cite{powerlaw}), capturing the intuition
that a few ingredients such as salt, sugar, and egg constitute a major part of
our every day diet. As a result, the set of distinct ingredients roughly follows
Heap's law, as seen in Fig.~\ref{heap_law}, with an exponent around $0.64$. According to the
method in paper~\cite{zipf}, the exponent of Zipf's law corresponding to
Fig.~\ref{degree_ingredient} can be estimated by $\frac{1}{\lambda-1}$. The
product of this exponent and the exponent of Heap's law (0.64) is close to 1,
which is consistent with the previous result~\cite{lvplos}.

\subsection{Quantifying similarity between cuisines}

Our dataset can be considered as a bipartite network with a set of recipes and
a set of ingredients. An edge between a recipe and an ingredient indicates that
the recipe contains the corresponding ingredient. Since each recipe belongs to
one and only one regional cuisine, the edges could be categorized into
cuisines. Given a cuisine $c$ and an ingredient $i$, we use  $n_{i}^{c}$ to denote
the degree of ingredient $i$, counted with edges in cuisine $c$. In other
words, $n_{i}^{c}$ is the number of recipes (in cuisine $c$) that use
ingredient $i$. Therefore, the ingredient usage vector of regional cuisine $c$
is written in the following form:
\begin{equation}
 \widetilde{P^{c}} = (p^{c}_{1},p^{c}_{2},\dots,p^{c}_{i},\dots,p^{c}_{n}),
\label{unweighted_p}
\end{equation}
where $p_{i}^{c} = \frac{n_{i}^{c}}{\sum_{i=1}{n_{i}^{c}}}$ is the fraction of
recipes in the cuisine $c$ that contain the ingredient $i$. For example, if
recipes in a regional cuisine $c$ use $1,000$ ingredients (with duplicates) in
total and ingredient $i$ appears in $10$ recipes in that cuisine, we have
$p^{c}_{i} = \frac{10}{1000}$.

Since common ingredients carry little information, we use an ingredient usage
vector inspired by TF-IDF (Term Frequency Inverse Document
Frequency)~\cite{tfidf}:
\begin{equation}
P^{c} = (w_{1}p^{c}_{1}, w_{2}p^{c}_{2},\dots, w_{j}p^{c}_{i},\dots,
w_{n}p^{c}_{n}),
\label{weighted_p}
\end{equation}
where a prior weight $w_{i} =
log\frac{\sum_{c}\sum_{i}{n_{i}^{c}}}{\sum_{c}{n_{i}^{c}}}$ is introduced to
penalize a popular ingredient. We use $P^{c}$ for all calculations in this
paper. With this representation in hand, we quantify the similarity between two
cuisines using the Pearson correlation coefficient (Eq.~\ref{PCC}) and cosine
similarity (Eq.~\ref{cos}).

(i) Pearson product-moment correlation \cite{PCC}: This metric measures the
extent to which a linear relationship is present between the two vectors. It is
defined as
\begin{equation}
PCC(P^{a},P^{b}) =
\frac{\sum_{i}{(P^{a}_{i}-\overline{P^{a}})(P^{b}_{i}-\overline{P^{b}})}}{\sqrt{\sum_{i}{(P^{a}_{i}-\overline{P^{a}})^{2}}}\sqrt{\sum_{i}{(P^{b}_{i}-\overline{P^{b}})^{2}}}},
\label{PCC}
\end{equation}
where $P^{a}$ and $P^{b}$ are ingredient usage vectors of
regional cuisine $a$ and $b$, respectively.

(ii) Cosine similarity \cite{cosine}: It is a measure of similarity between two
vectors of an inner product space that measures the cosine of the angle between
them. For regional cuisines $a$ and $b$,  the cosine similarity is represented
using a dot product and magnitude as
\begin{equation}
cosine(P^{a},P^{b}) = \frac{\sum_{i}{P^{a}_{i}P^{b}_{i}}}{\sqrt{\sum_{i}{(P^{a}_{i})^{2}}}
    \times \sqrt{\sum_{i}{(P^{b}_{i})^{2}}}}.
\label{cos}
\end{equation}


\section{Results}

\subsection{Outlier detection}

\begin{figure*}[htbp]
  \centering
  \includegraphics[width = 7.3cm,height=6cm]{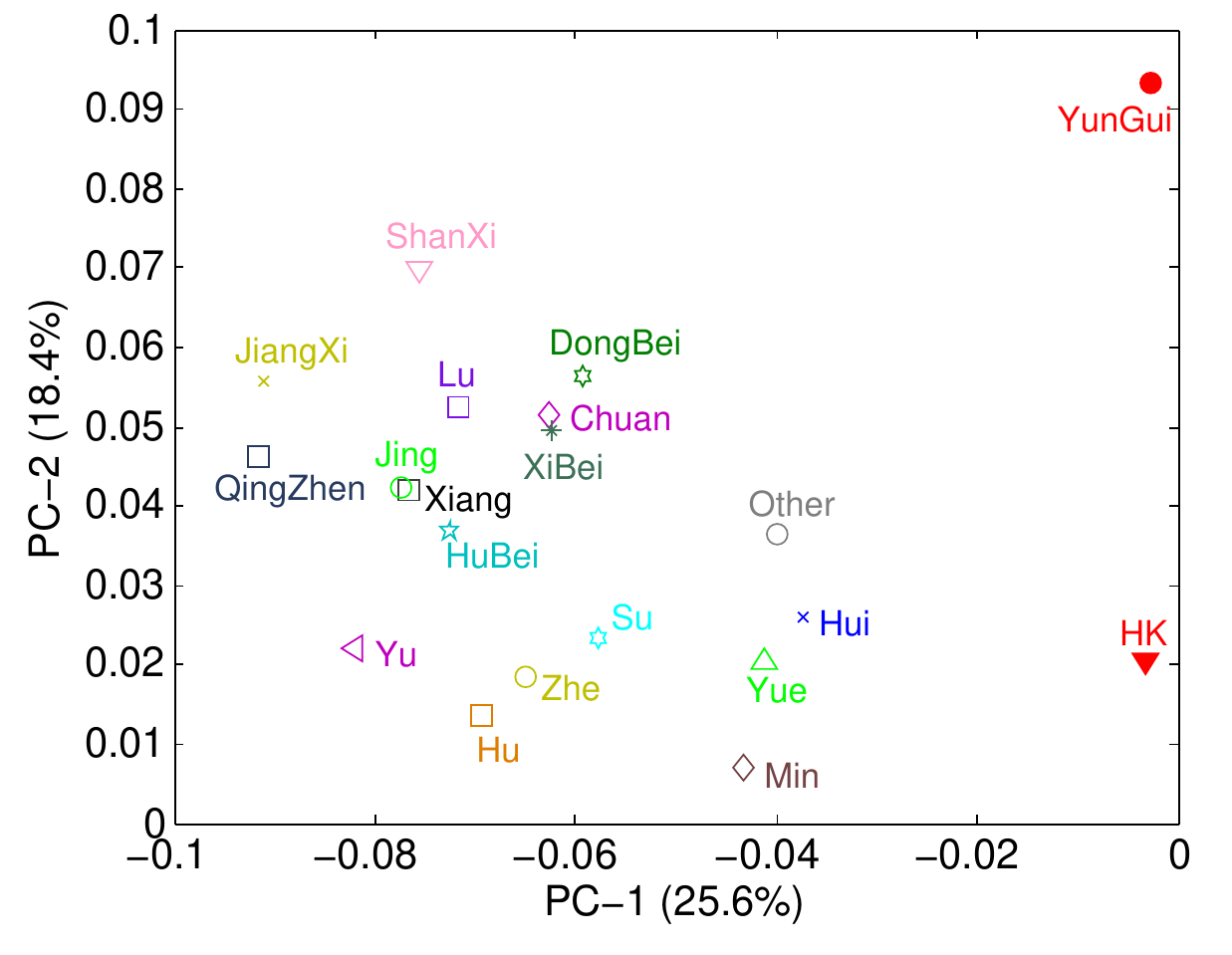}
  \caption{ {\bf The distribution of regional cuisines in the first two principal component spaces.}
  YunGui and Hong Kong (red) stand out as outliers. \label{2d}}
\end{figure*}

As an overview, we apply Principal Component Analysis to identify principal components of the
ingredient usage matrix~\cite{PCA}. The distributions of regional
cuisines in two principal components (capturing 44\% of the information)
are presented in Fig.~\ref{2d}. The two most obvious outliers (in solid red)
are YunGui and Hong Kong cuisine. This may reflect the facts that ethnic minorities have historically resided in the YunGui region and that Hong Kong was ruled by the British Empire
and Japan for more than 100 years.

\subsection{Geography and climate}

Climate shapes ingredient usage patterns not only by affecting the availability
of ingredients but also by exerting other pressures, such as a need to use
additional spices as preservatives~\cite{spices}. At the
same time, we expect nearby regions to have a higher probability of  similar
food culture even without similar climate, because they are more likely to
have more communication and migration.

To estimate the effect of climate we use temperature
as a proxy. We assume that the annual average
temperature approximately captures one of the most fundamental
aspects of climates. As shown in the previous work on
spices~\cite{spices}, annual temperature strongly predicts the usage of
spices, and we further assume that temperature is a strong climate factor that affects
ingredient availability. For two regions $A$ and $B$, the temperature
difference $\Delta T_{A,B}$ is simply $|T_{A} - T_{B}|$, where $T_{A}$ is the
annual average temperature of region $A$.

\begin{figure*}[htbp]
  \centering
  \includegraphics[width = 7.3cm,height=6cm]{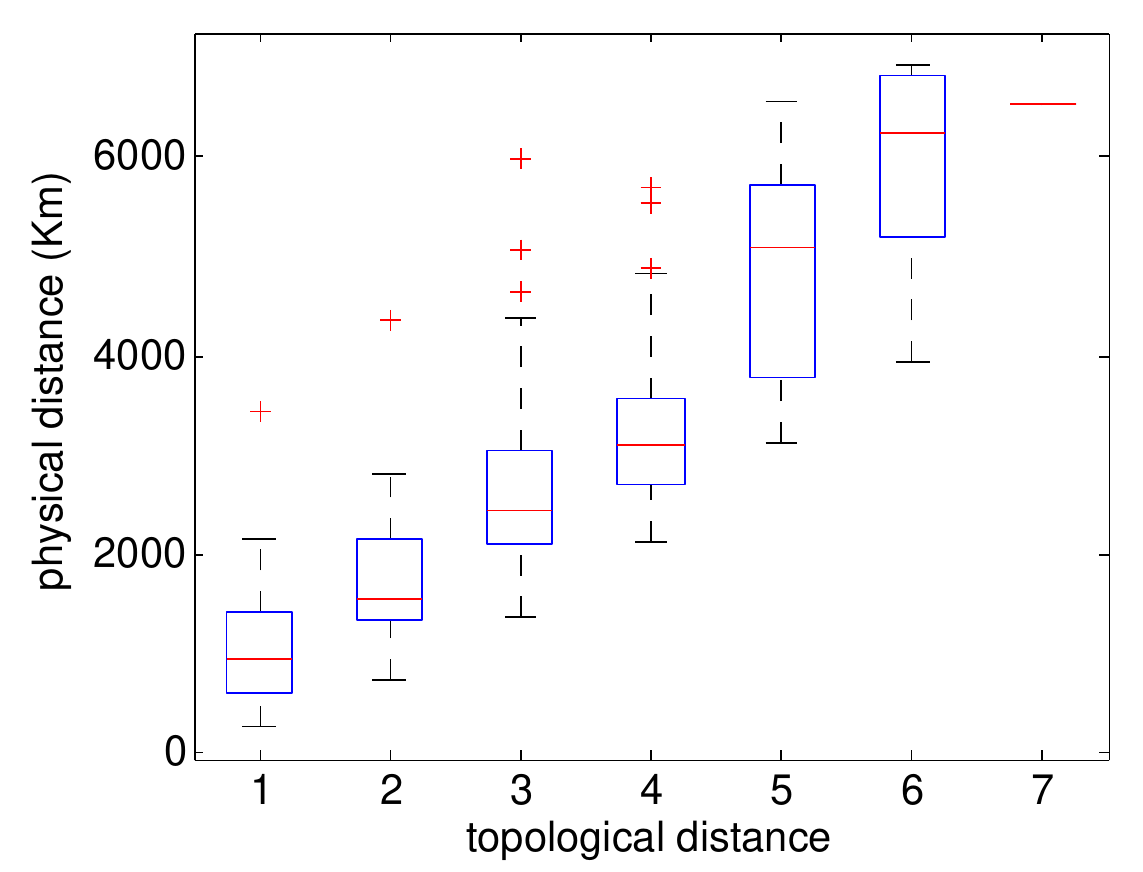}\\
  \caption{ {\bf Topological distance vs. physical distance.} }\label{twodistance}
\end{figure*}

We quantify geographical proximity using two distance measures: physical
distance and topological distance. We measure physical distance between two
cuisines by identifying the central cities of the cuisines and then calculating
the great-circle distance~\cite{sphericaldistance}. To measure topological
distance between two cuisines, we construct a graph of
cuisines, where a node represent a regional cuisine and an edge represents
adjacency of two cuisines, we then measure topological distance by
the shortest path length on the graph.
Figure~\ref{twodistance} shows that the geographical distance and topological
distance are correlated yet exhibit large variance.

\begin{figure*}[htbp]
  \centering
  \includegraphics[width = 7.2cm,height=5.8cm]{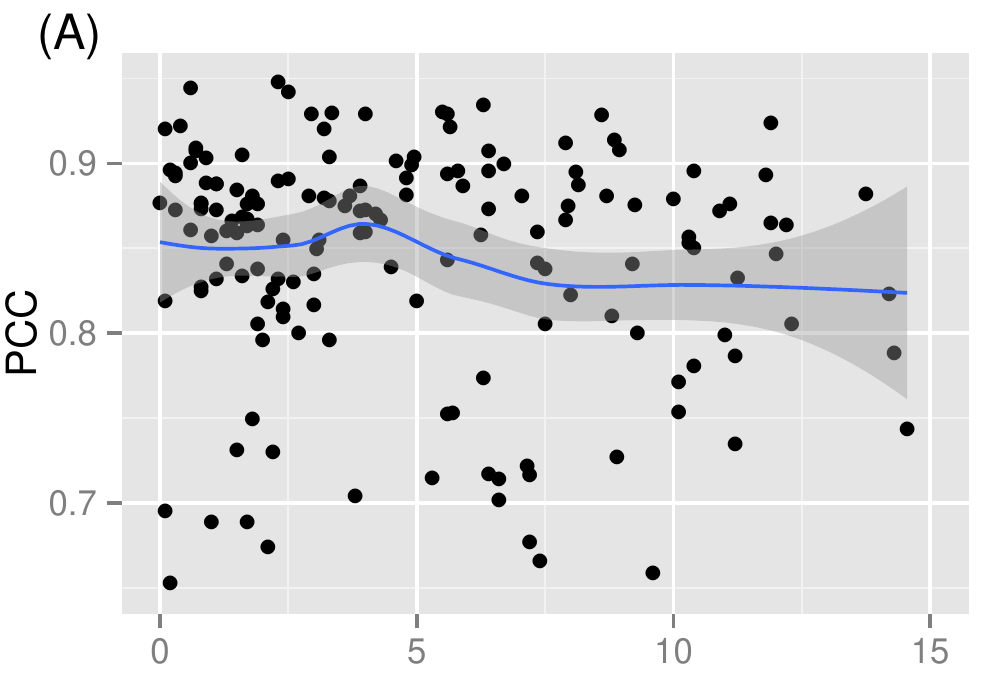}
  \includegraphics[width = 7.2cm,height=5.8cm]{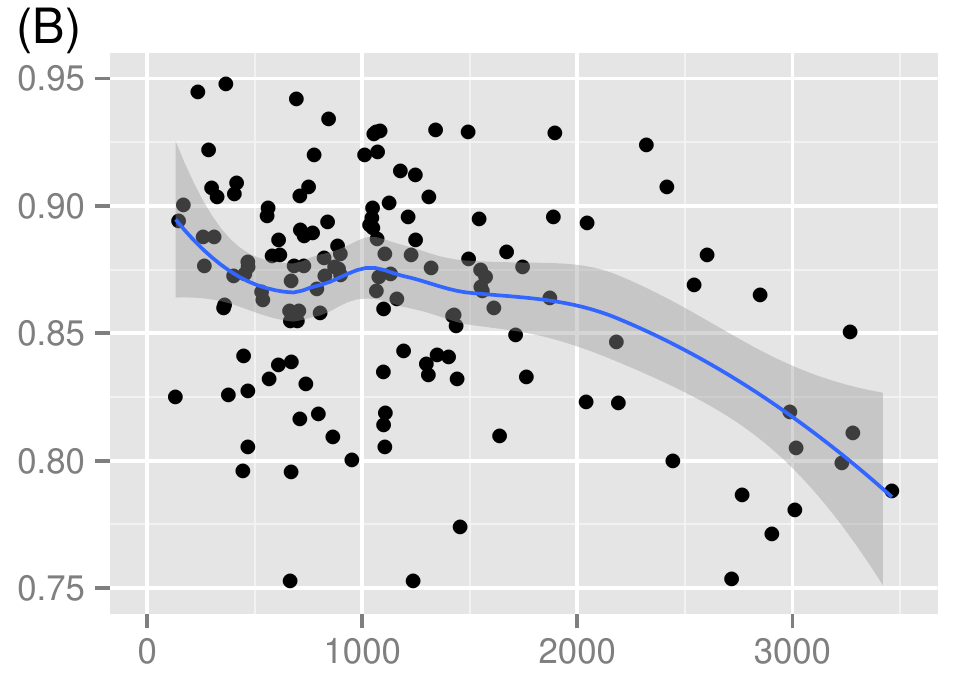}
  \includegraphics[width = 7.2cm,height=6cm]{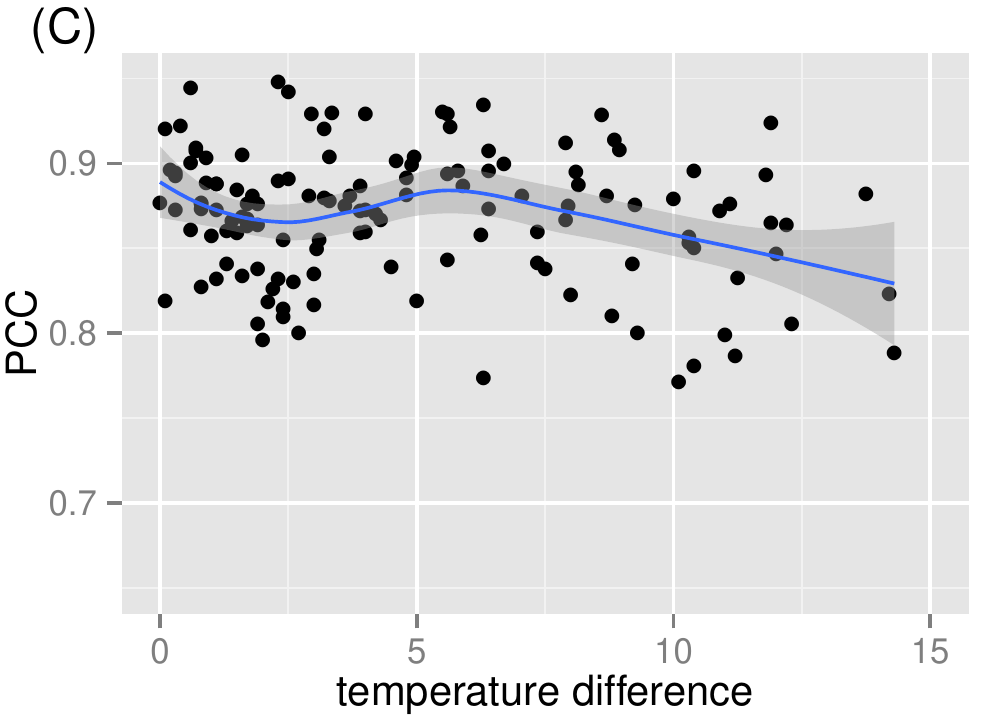}
  \includegraphics[width = 7.2cm,height=6cm]{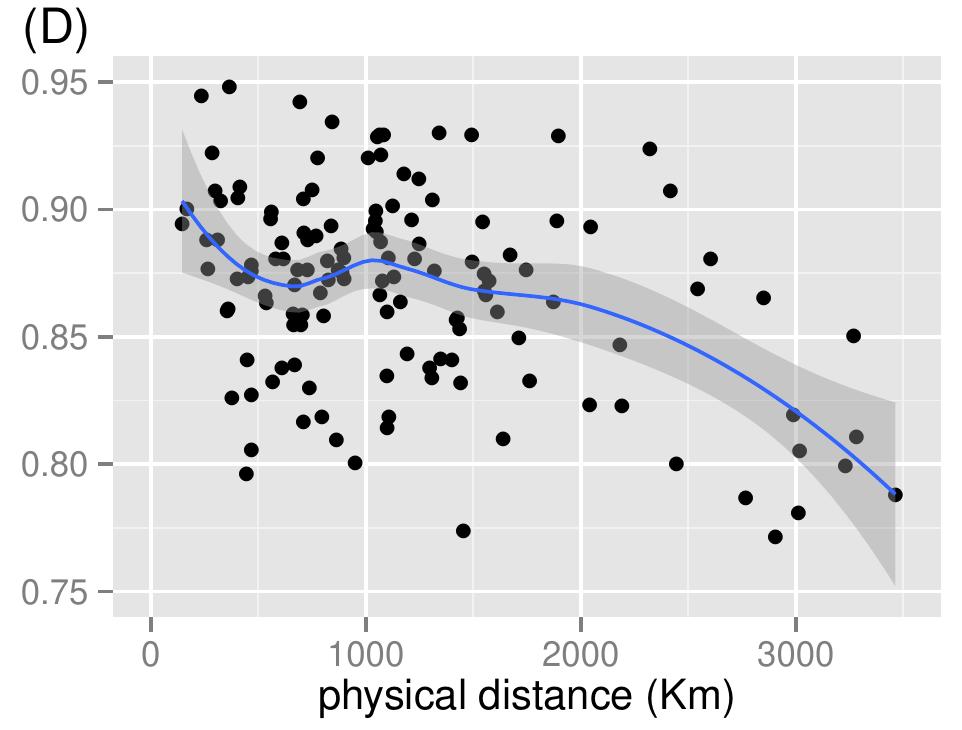}
  \caption{ {\bf The dependence of similarities between different regional cuisines on the climate and geography.}
  (A): scatter plot of PCC and temperature difference (all regional cuisine pairs);
  (B): scatter plot of PCC and physical distance (all regional cuisine pairs);
  (C): scatter plot of PCC and temperature difference (neglecting outliers);
  (D): scatter plot of PCC and temperature difference (neglecting outliers); } \label{climate_geodis_sim}
\end{figure*}

Figure~\ref{climate_geodis_sim} compares how temperature and physical distance
are related to the similarities between regional cuisines. The left
column shows the results of temperature and the right column
shows that of physical distance. Each circle represents a pair of
cuisines. The Pearson correlation coefficient between the temperature difference and PCC
is -0.134 (Figure~\ref{climate_geodis_sim}A), indicating a weak correlation
between similarity of regional cuisines and their temperature
difference. When we delete the two outliers mentioned above (YunGui and
HK), the Pearson correlation coefficient between the temperature difference and
PCC becomes -0.216 (Figure~\ref{climate_geodis_sim}C). That is, regions with
similar temperature tend to share similar usage patterns of
ingredients, which is consistent with  previous results~\cite{spices}.
However, this may not be the effect of temperature, because climate is
correlated with distance. The Pearson correlation coefficient
between the physical distance and PCC is -0.289
(Figure~\ref{climate_geodis_sim}B), indicating a stronger correlation. When
neglecting outliers, it becomes -0.385 (Figure~\ref{climate_geodis_sim}D).
The $p$-values of all cases indicate significant difference
($p\ll0.05$ for both cases).

The previous analysis does not provide a complete picture, since
geography and climate are strongly correlated; nearby regions are more likely
to have similar climates. To estimate the effect of climate and geographical
proximity, we calculate partial correlation~\cite{partialcorreation}, which
is used to measure the linear association between two factors while
removing the effect of other additional factors.
The partial correlation coefficients between physical distance and PCC, given
temperature difference as a control variable, is -0.280. However,
the partial correlation between temperature difference and
ingredient usage similarity, given physical distance as a control
variable, the expected negative correlation completely vanishes and
the correlation coefficient becomes 0.116.
Our results indicate that the effect
of temperature on the ingredient usage pattern may not exist at all. The
results with cosine and the cases without outliers also show the same
tendency.


\begin{figure*}[htbp]
\centering
 \includegraphics[width = 7.3cm,height=6cm]{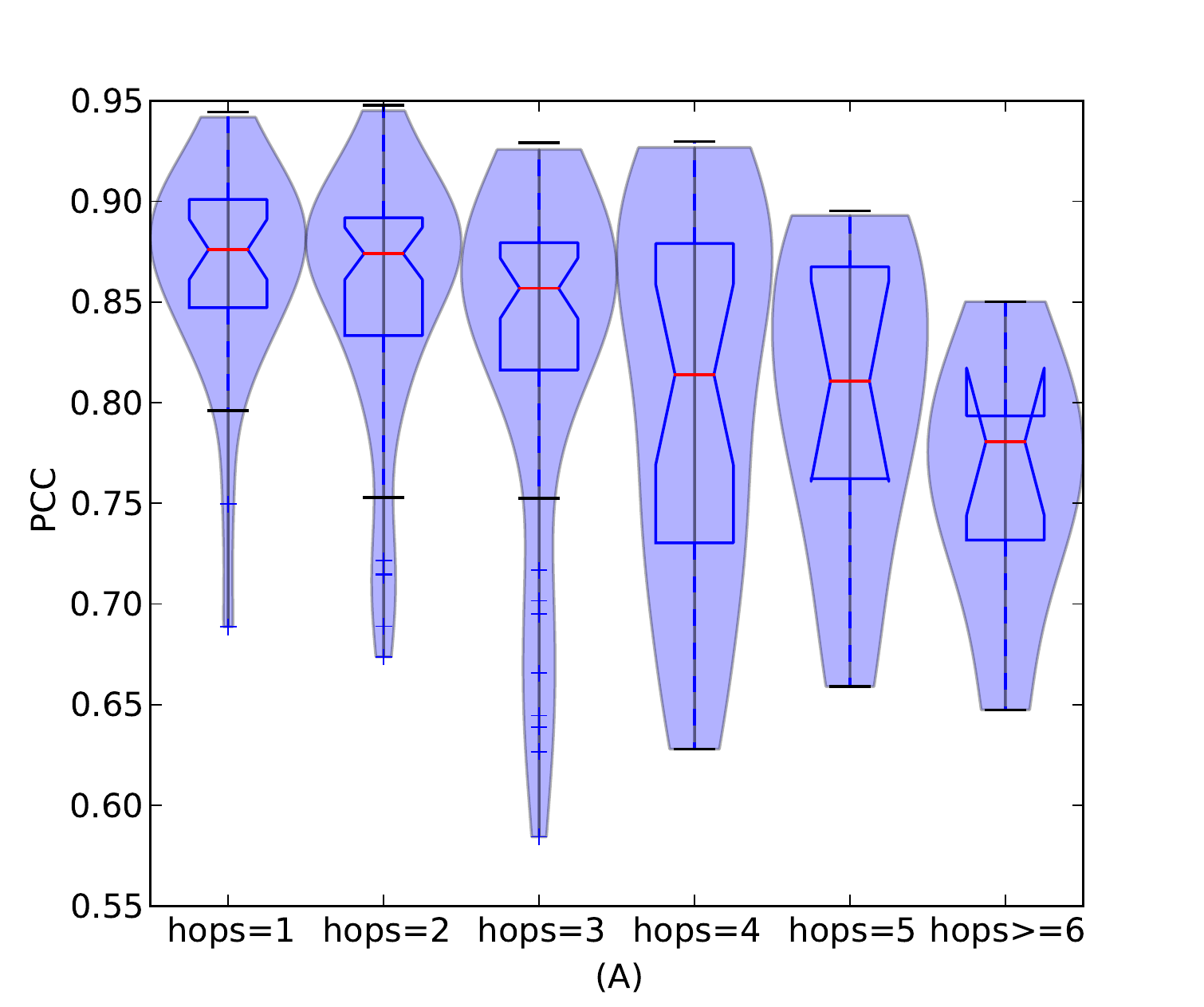}
 \includegraphics[width = 7.3cm,height=6cm]{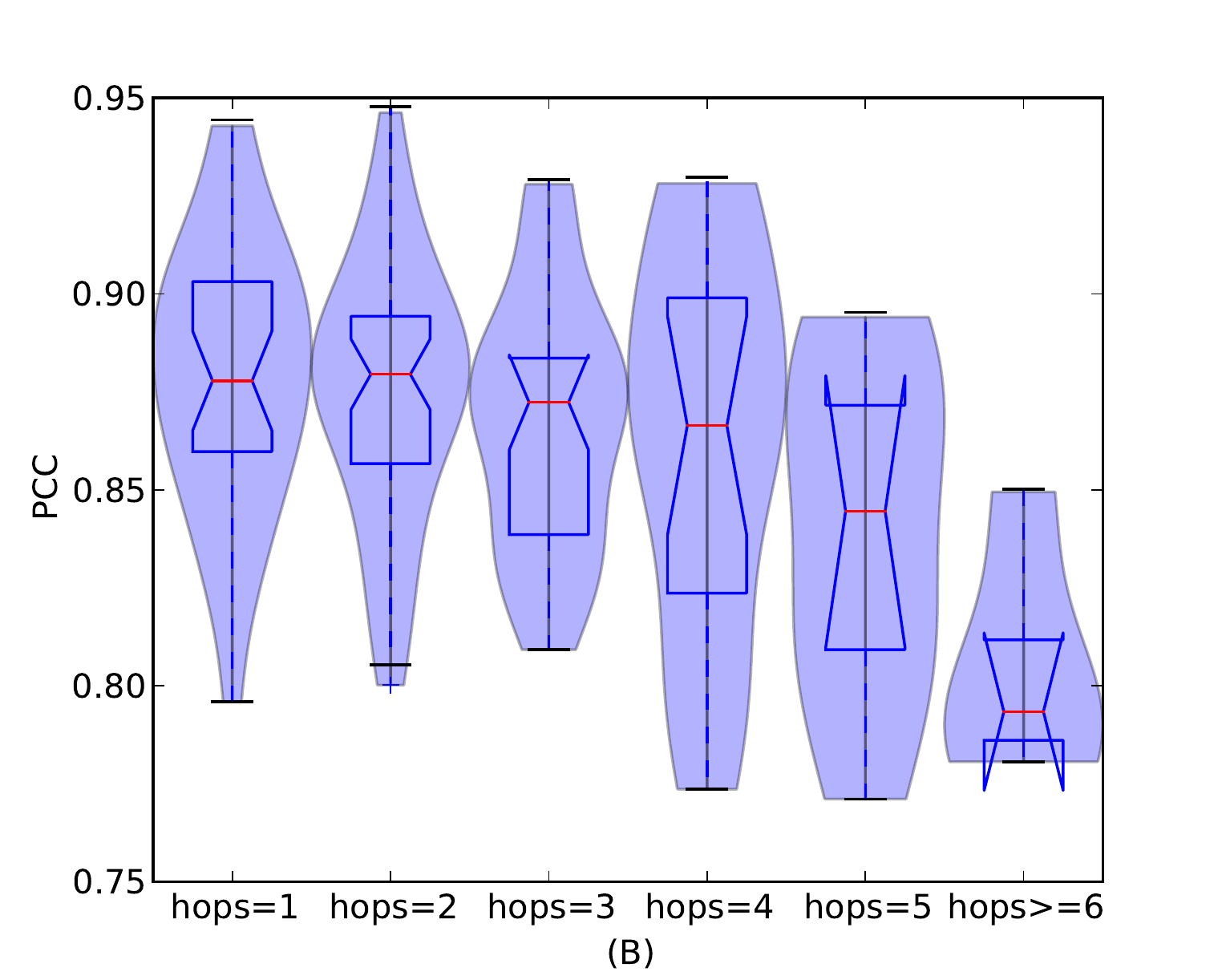}
\caption{ {\bf Similarity distributions of regional cuisine pairs with different topological distance ($hops=1$, $hops=2$, $\dots$, $hops\geq6$).} (A): result for all regional cuisines; (B): result when neglecting outliers.\label{violin_ing}}
\end{figure*}

Here we examine the relationship between topological distance and the similarity of cuisines.
Figure~\ref{violin_ing} shows the similarity distribution of cuisines with
respect to topological distance. Analysis of variance (ANOVA)~\cite{ANOVA}
shows that the difference in the similarity distribution is significant
($p\ll0.001$ for both cases).
The figure shows a clear trend that geographically closer
regional cuisines have more similar ingredient usage patterns.
We perform a simple permutation significance test by classifying
all regional cuisine pairs into two classes: the first class contains regional
cuisine pairs with the topological distance less than or equal to 2 and the
second class contains those of topological distance larger than 2.
Denote the similarities of two classes
as $F$ (far, with topological distance $>2$) and $C$ (close, with topological
distance $\leq2$), and $\overline{F}$ and $\overline{C}$ is the mean of $F$ and
$C$, respectively. $n_{F}$ and $n_{C}$ are the sample sizes corresponding to each
group. The null hypothesis $H_{0}$ says the two classes $F$ and $C$ have
identical probability distributions. We performed the test as follows. First,
the difference between $F$ and $C$ is calculated. This is the observed value of the
test statistic, namely $T_{obs} = \overline{C} - \overline{F}$. The observations of classes $F$ and $C$ are then pooled. Next, the difference in
sample means is calculated and recorded for every possible way of dividing
these pooled values into two groups of size $n_{F}$ and $n_{C}$, denoted by
$T$. Lastly the one-sided $p$-value of the test is calculated as $p = p (T >
T_{obs})$. The $p$-values of Fig.~\ref{violin_ing} indicate significant difference
($p\ll0.01$ for both cases).

\section{Model and Validation}

We build an evolution model of Chinese cuisines based on the simple notion
that geographical proximity breeds more communication and migration. Our model
uses the copy-and-mutate model of recipe evolution~\cite{osameNJP}. We tested
two models that use topological distance and physical distance respectively,
and found that the model using topological distance produces better results.
There are three elements
in the model: regional cuisines, recipes, and ingredients. We assume the
same set of Chinese regional cuisines in the same location. A recipe is a set
of ingredients and belongs to one of the regional cuisines. We assume that each
recipe has exactly the same number of ingredients ($K$). Each ingredient $i$ has
a fitness value $f_{i}$, randomly drawn from a uniformly distribution in
$[0,1]$. This fitness represents intrinsic properties such as nutritional
value, flavor, cost, and availability~\cite{osameNJP}. All
the symbols used in our model are listed in Table~\ref{symbol}.

\begin{table*}[!ht]
\caption{Notations for parameters and quantities in the model}
\centering
{\begin{tabular}{c|l}
  \hline\hline
$N_{i}$& number of all ingredients\\
$N_{i}(t)$& number of initial ingredients at time $t$\\
$N_{r}$& number of all recipes \\
$N_{r}^{c}$ &number of recipes in the regional cuisine $c$ \\
$N_{r}^{c}(t)$ &number of initial recipes in the regional cuisine $c$ at time $t$\\
$K$& number of ingredients per recipe\\
$K'$& number of ingredients to be mutated in each recipe, $K'\leq K$ \\
$P_{L}$&probability of interaction\\
$f_{i}$& the fitness of ingredient $i$\\
$L_{i,j}$ & the topological distance of regional cuisine $i$ and $j$ \\
\hline\hline
\end{tabular}\label{symbol}}
\end{table*}

Let us describe our model in detail. In the initial state there are
$N_{i}(0)=20$ ingredients in the ingredient pool, and each regional cuisine
contains one recipe that consists of $K$ random ingredients chosen from the
initial ingredient set; that is, $N_{r}^{c}(0)=1$.

\noindent\textbf{Step 1}: Choose one regional cuisine preferentially. Regional
cuisine $c$ is chosen with probability \begin{equation} p(c) =
\frac{N_{r}^{c}}{N_{r}}, \end{equation} where $N_{r}^{c}$ is the number of
recipes of regional cuisine $c$ and $N_{r}$ is the total number of recipes of
all regional cuisines. With probability $P_{L}$ the chosen regional cuisine
($c$) will interact with another region (\noindent\textbf{Step 2}).
With probability $1-P_{L}$
cuisine $c$ will develop a recipe itself by randomly selecting $K$
unique ingredients from the ingredient pool.

\noindent\textbf{Step 2}: Generate a recipe by learning from others. With probability proportion
to their topological distance, cuisine $c$ will select another regional cuisine
$c'$. Regional cuisine $c'$ will be chosen with probability \begin{equation}
p(c') = \frac{ (1+L_{c,c'})^\alpha }{\sum_{c'}(1+L_{c,c'})^\alpha},
\end{equation} where $L_{c,c'}$ is the topological distance between cuisine $c$
and $c'$. $\alpha$ is a free parameter to tune the importance of
topological distance. Regional cuisines
with smaller topological distance from cuisine $c$ will have higher probability
to be chosen in this step when $\alpha<0$. Assuming cuisine $c'$ is chosen to
interact with $c$, then cuisine $c$ will randomly choose one recipe (as a
template) of cuisine $c'$ as a template and copy it. From this copy, we
randomly chose an ingredient $i$ and compare it
with an ingredient $j$ that is randomly chosen from the ingredients pool, if
$f_{j}>f_{i}$, we replace ingredient $i$ by ingredient $j$. This process is
repeated $K'$ times. We then execute \noindent\textbf{Step 3}.

\noindent\textbf{Step 3}. Add new ingredients. With probability
$\frac{N_{i}- N_{i}(t)}{N_{r}-N_{r}(t)}$ we add one new ingredient to the
ingredients pool and replace the ingredient having the smallest
fitness in the recipe that was generated in the previous steps with
the new ingredient in order to assure all the ingredients in the
ingredient pool have been used. We then add the modified new recipe to the pool
of cuisine $c$. If no new ingredient is added in this step, then add the new recipe to the
pool of cuisine $c$ without modification. If we already have  $N_{r}$
recipes, stop the simulation, or else repeat the previous steps.

\begin{figure*}[htbp]
  \centering
  \includegraphics[width = 7.5cm,height=6cm]{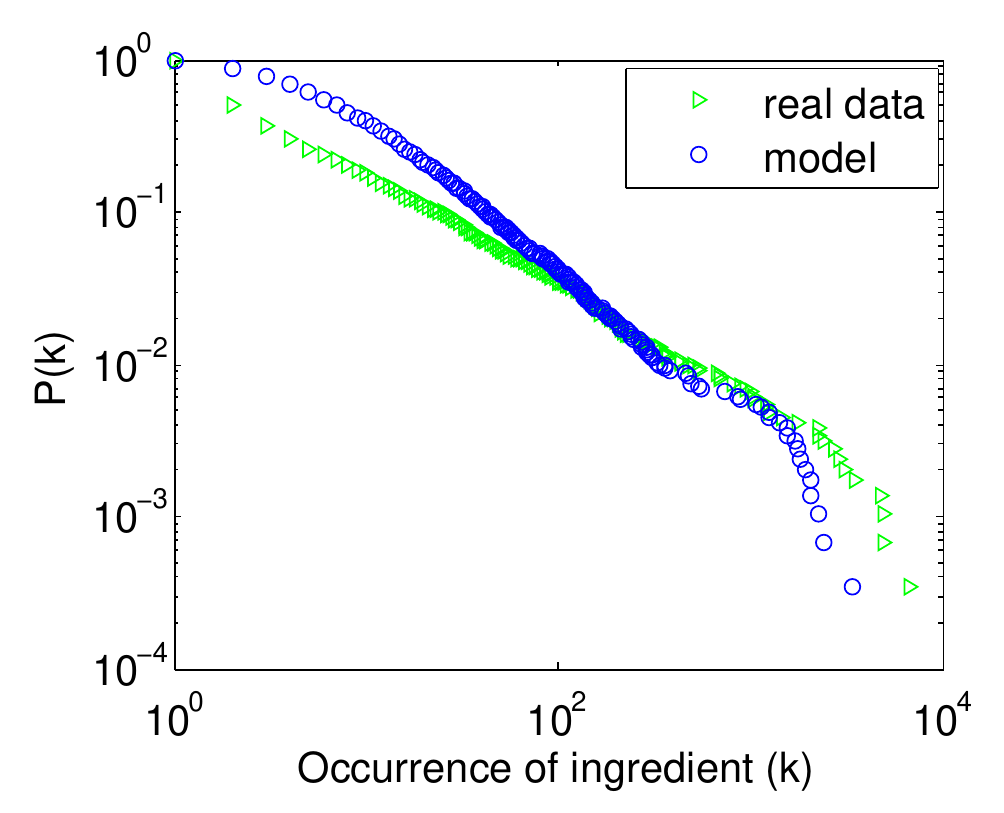}
  \caption{{\bf Cumulative frequency distribution of the cuisine network generated by our model compared to the the empirical distribution from the dataset of Chinese cuisine.} \label{degree_model}}
\end{figure*}

The parameters $N_{r}^{c}$ and $N_{i}$ are obtained from the data.
For simplicity, we set $K=10$, which means that every recipe has
ten unique ingredients. In the initial state, there are $N_{i}(0)=20$
ingredients in the ingredient pool, and each regional cuisine contains one
recipe that choose randomly from the initial ingredients; that is,
$N_{r}^{c}(0)=1$. The cuisines then evolve as new recipes and ingredients join
over time. We used $k'=2$, $P_{learn}=0.85$, and $\alpha=-5$, which generates
results that closely resemble the empirical findings.
Figure~\ref{degree_model} demonstrates that our model produces a qualitatively
similar, skewed degree distribution.

\begin{figure*}[htbp]
  \centering
  \includegraphics[width = 7.3cm,height=6cm]{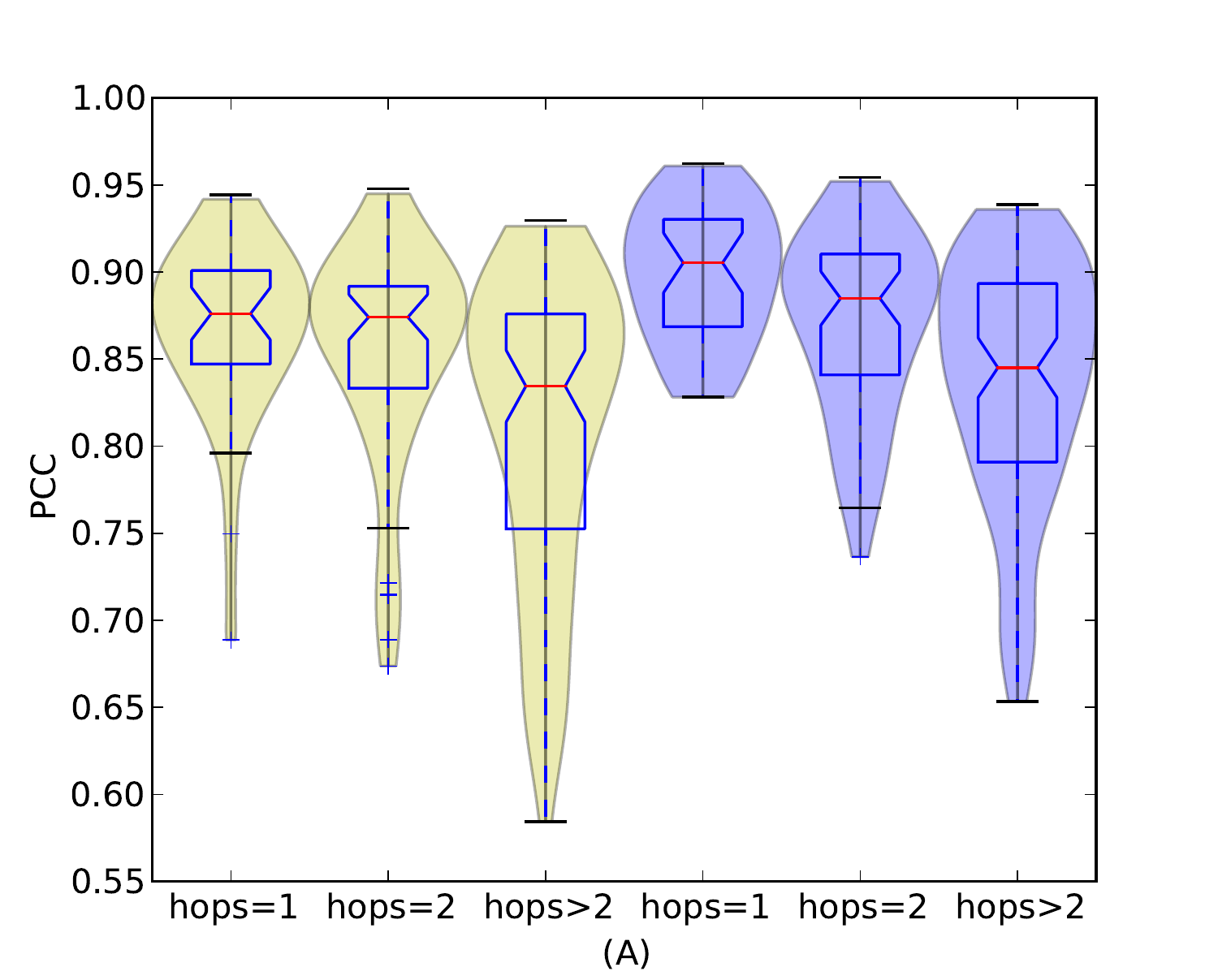}
  \includegraphics[width = 7.3cm,height=6cm]{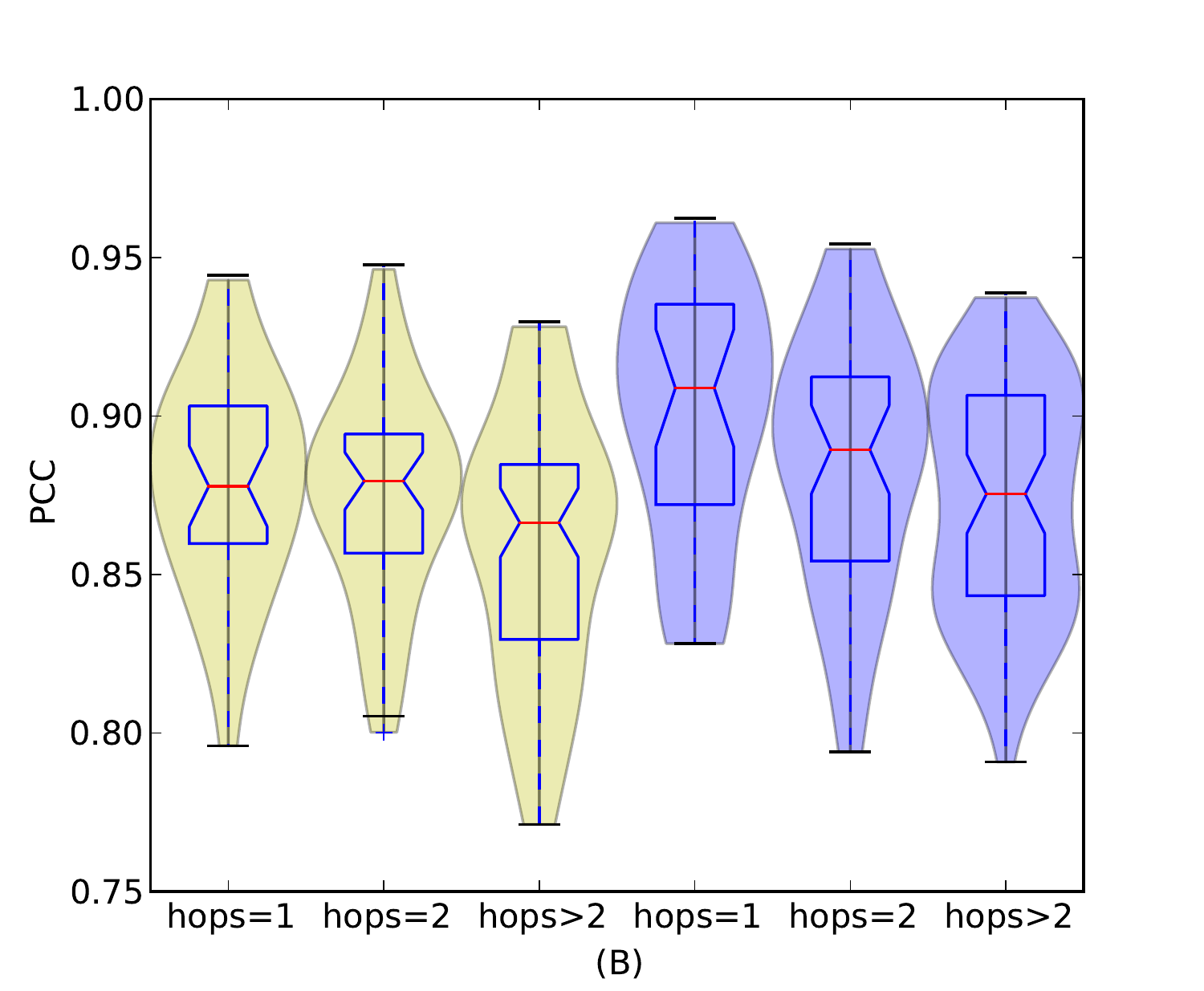}
  \caption{ {\bf Similarity distributions of regional cuisine pairs with different topological distance ($hops=1$, $hops=2$ and $hops>2$).} The blue ones are the results generated by our model, while the yellow ones are results of the dataset. (A): result for all regional cuisines; (B): result when neglecting outliers.\label{violin_model}}
\end{figure*}

\begin{figure*}[htbp]
  \centering
  \includegraphics[width = 7.3cm,height=6.1cm]{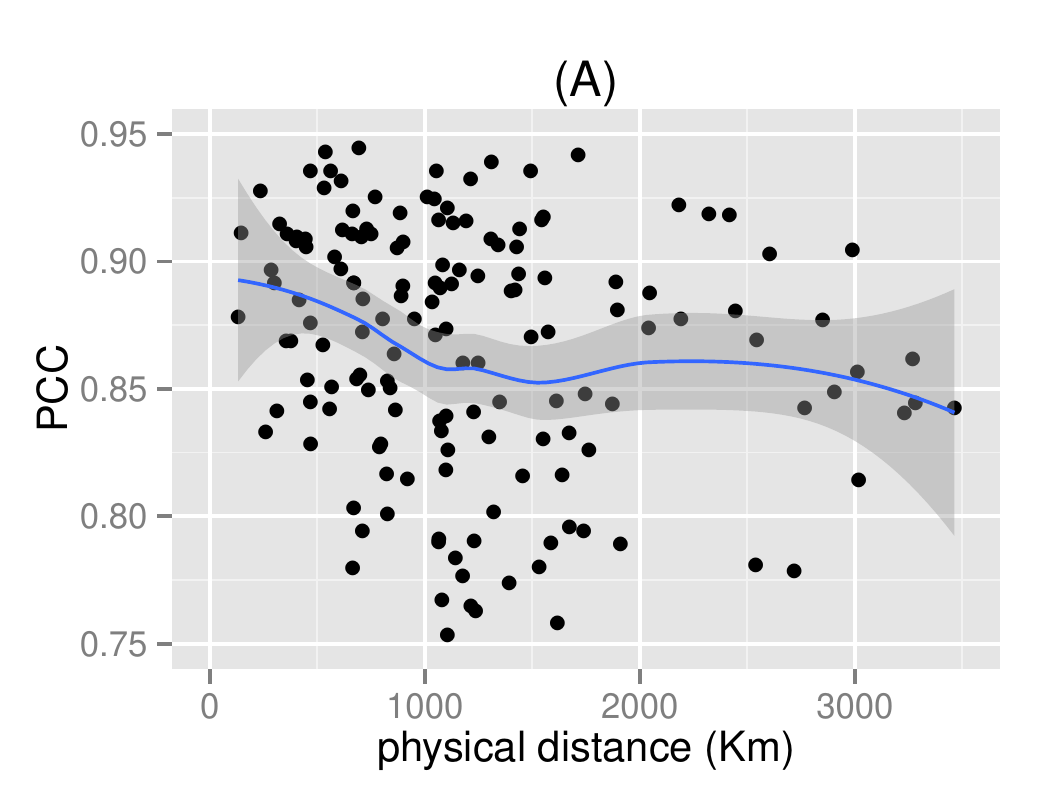}
  \includegraphics[width = 7.3cm,height=6.1cm]{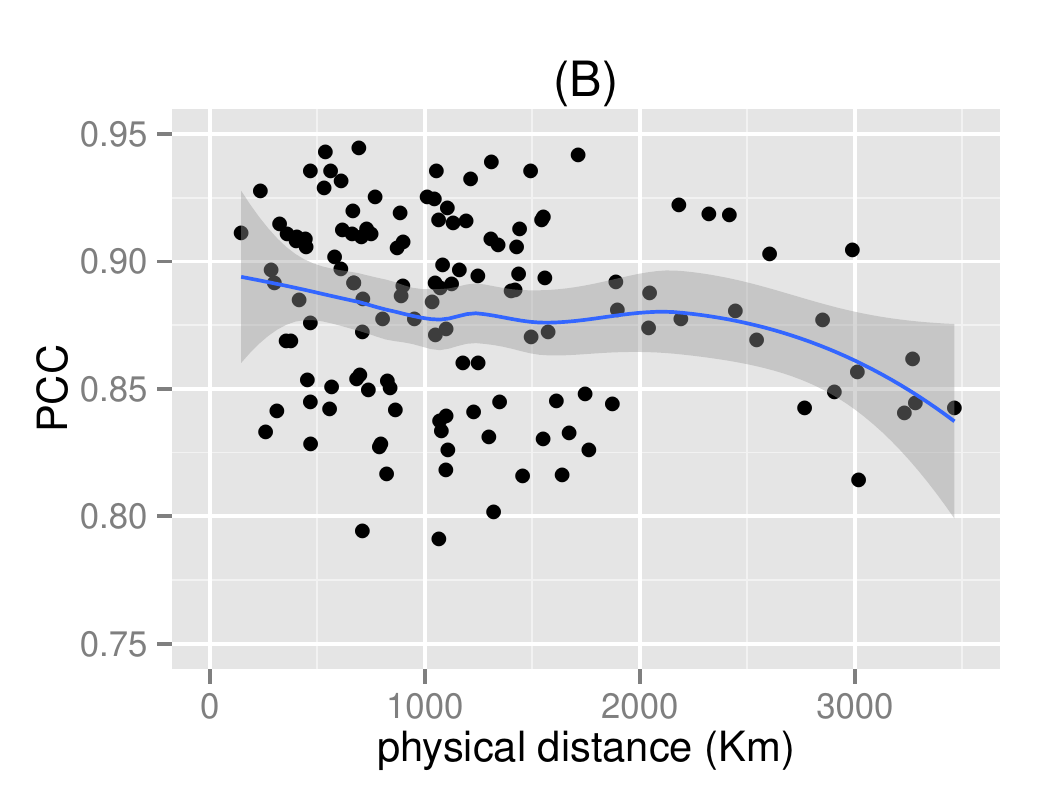}
  \caption{ {\bf The dependence of similarities between different regional cuisines on the geographical distance in the model.} (A): result for all regional cuisines; (B): result when neglecting outliers.\label{sim_model2}}
\end{figure*}

Figure~\ref{violin_model} shows the comparison between the real data and our model in terms of the
similarity distribution with topological distance.
Fig.~\ref{violin_model}A is the result of all regional
cuisine pairs, displaying that our model can achieve the
similar tendency as the real dataset, although the
real dataset shows more diversity than our model.
We think this is a result of the existance of outliers in the dataset.
Fig.~\ref{violin_model}B shows the result without outliers,
displaying a better match with our dataset.
Figure~\ref{sim_model2} shows the
dependency between physical distance and cuisine similarity in our model.
The results display a similar tendency
found in the real data (C and D in Fig.~\ref{climate_geodis_sim}).
Results of cosine similarity
are not displayed, as all of them display similar tendencies with PCC.

\section{Conclusion and Discussion}

We empirically analyzed the similarity relations between major regional
cuisines in China in terms of climate and physical proximity.
We find that climate (temperature) does not show any correlation with ingredient usage
similarity if we control geographical distance, while geographical proximity
seems to be a key factor in the shaping of regional cuisines.
Based on the finding, we propose a copy-and-mutate model that
incorporates geographical proximity. We show that the results of our model
agree with our empirical findings.

\section*{Acknowledgments}

We thanks Huawei Shen, Yu Huang, Shaojian Wang and Giovanni Luca Ciampaglia for helpful discussion
and useful suggestion, and John McCurley for editorial
assistance.
This work is partially supported by the National Natural Science
Foundation of China under Grant Nos. 11105024, 11205042 and 11105025. YXZ
acknowledges the supporting from the Program of Outstanding PhD Candidate in
Academic Research by UESTC (Grant No. YBXSZC20131035) and China Scholarship
Council. QMZ acknowledges the supporting from the Program of Outstanding PhD
Candidate in Academic Research by UESTC (Grant No. YBXSZC20131034).

\bibliography{bibfile}

\begin{thebibliography}{10}
\providecommand{\url}[1]{\texttt{#1}}
\providecommand{\urlprefix}{URL }
\expandafter\ifx\csname urlstyle\endcsname\relax
  \providecommand{\doi}[1]{doi:\discretionary{}{}{}#1}\else
  \providecommand{\doi}{doi:\discretionary{}{}{}\begingroup
  \urlstyle{rm}\Url}\fi
\providecommand{\bibAnnoteFile}[1]{%
  \IfFileExists{#1}{\begin{quotation}\noindent\textsc{Key:} #1\\
  \textsc{Annotation:}\ \input{#1}\end{quotation}}{}}
\providecommand{\bibAnnote}[2]{%
  \begin{quotation}\noindent\textsc{Key:} #1\\
  \textsc{Annotation:}\ #2\end{quotation}}
\providecommand{\eprint}[2][]{\url{#2}}

\bibitem{foodandcuture}
Counihan C, Esterik PV (1997) Food and culture: a reader.
\newblock Routledge, London.
\bibAnnoteFile{foodandcuture}

\bibitem{audrey1932}
Richards AI (1939) Hunger and work in a savage tribe.
\newblock Routledge, London.
\bibAnnoteFile{audrey1932}

\bibitem{raymond1934}
Firth R (1934) The sociological study of native diet.
\newblock Journal of the international institute of african languages and
  cultures VII: 401-414.
\bibAnnoteFile{raymond1934}

\bibitem{history1}
Moyers S (1996) Garlic in health, history, and world cuisine.
\newblock Suncoast Press.
\bibAnnoteFile{history1}

\bibitem{history2}
Tregear A (2003) From Stilton to Vimto: Using Food History to Re-think Typical
  Products in Rural Development.
\newblock Wiley-Blackwell.
\bibAnnoteFile{history2}

\bibitem{history3}
Civitello L (2011) Cuisine and Culture: A History of Food and People.
\newblock Wiley; 3th edition.
\bibAnnoteFile{history3}

\bibitem{history4}
Diamond JM (1999) Guns, Germs, and Steel: The Fates of Human Societies.
\newblock W. W. Norton and Company.
\bibAnnoteFile{history4}

\bibitem{sociology1}
Mennell SJ, Murcott A, Otterloo AHv (1993) The Sociology of Food: Eating, Diet
  and Culture.
\newblock Sage Publications; 2nd edition.
\bibAnnoteFile{sociology1}

\bibitem{sociology2}
Beardsworth A, Keil T (1997) Sociology on the Menu: An Invitation to the Study
  of Food and Society.
\newblock Routledge.
\bibAnnoteFile{sociology2}

\bibitem{sociology3}
Germov J, Williams L (2004) A Sociology of Food and Nutrition: The Social
  Appetite.
\newblock Oxford University Press.
\bibAnnoteFile{sociology3}

\bibitem{philosophy1}
Allhoff F, Monroe D (2007) Food and Philosophy: Eat, Think, and Be Merry.
\newblock Wiley-Blackwell; 1st edition.
\bibAnnoteFile{philosophy1}

\bibitem{philosophy2}
Curtin DW, Heldke LM (1992) Cooking, Eating, Thinking: Transformative
  Philosophies of Food.
\newblock Indiana University Press; 1st Edition edition.
\bibAnnoteFile{philosophy2}

\bibitem{literary1}
Skubal SM (2002) Word of Mouth: Food and Fiction After Freud.
\newblock Routledge.
\bibAnnoteFile{literary1}

\bibitem{spices}
Sherman PW, Billing J (1999) Darwinian gastronomy: Why we use spices.
\newblock Bioscience 49: 453-463.
\bibAnnoteFile{spices}

\bibitem{YY1}
Ahn YY, Ahnert SE, Bagrow JP, Barab\'asi AL (2011) Flavor network and the
  principles of food pairing.
\newblock Scientific Reports 1: srep00196.
\bibAnnoteFile{YY1}

\bibitem{osameNJP}
Kinouchi O, Diez-Garcia RW, Holanda AJ, Zambianchi P, Roque AC (2008) The
  non-equilibrium nature of culinary evolution.
\newblock New Journal of Physics 10: 073020.
\bibAnnoteFile{osameNJP}

\bibitem{YY2}
Ahn YY, Ahnert SE (2013) The flavor network.
\newblock Leonardo 46: 272-273.
\bibAnnoteFile{YY2}

\bibitem{powerlaw-barabasi}
Barab\'asi AL, Albert R (1999) Emergence of scaling in random networks.
\newblock Science 286: 509-512.
\bibAnnoteFile{powerlaw-barabasi}

\bibitem{powerlaw}
Clauset A, Shalizi CR, Newman M (2009) Power-law distributions in empirical
  data.
\newblock SIAM Review 51: 661-703.
\bibAnnoteFile{powerlaw}

\bibitem{zipf}
Adamic LA (2000) Zipf, power-law, and pareto - a ranking tutorial.
\bibAnnoteFile{zipf}

\bibitem{lvplos}
L\"u L, Zhang ZK, Zhou T (2010) Zipf's law leads to heaps' law: Analyzing their
  relation in finite-size systems.
\newblock PLoS ONE 5: e14139.
\bibAnnoteFile{lvplos}

\bibitem{tfidf}
Salton G, McGill MJ (1986) Introduction to Modern Information Retrieval.
\newblock McGraw-Hill Book Company, New York.
\bibAnnoteFile{tfidf}

\bibitem{PCC}
Rodgers JL, Nicewander AW (1988) Thirteen ways to look at the correlation
  coefficient.
\newblock The American Statistician 42: 59-66.
\bibAnnoteFile{PCC}

\bibitem{cosine}
Tan PN, Steinbach M, Kumar V (2005) Introduction to Data Mining.
\newblock Addison-Wesley.
\bibAnnoteFile{cosine}

\bibitem{PCA}
Jolliffe IT (2008) Principal Component Analysis.
\newblock Springer, New York.
\bibAnnoteFile{PCA}

\bibitem{sphericaldistance}
http://mathworld.wolfram.com/SphericalDistance.html.
\newblock [Online; accessed 12-June-2013].
\bibAnnoteFile{sphericaldistance}

\bibitem{partialcorreation}
Cram\'er H (1999) Mathematical Methods of Statistics.
\newblock Princeton University Press.
\bibAnnoteFile{partialcorreation}

\bibitem{ANOVA}
Freedman DA (2009) Statistical Models: Theory and Practice.
\newblock Cambridge University Press.
\bibAnnoteFile{ANOVA}

\end{thebibliography}



\end{document}